\documentclass[12pt]{book}

\usepackage[dvips]{graphicx,color}
\usepackage{makeidx,tsukuba}

\makeauthorindex
\makeindex

\begin{document}

\BookTitle{\itshape The 28th International Cosmic Ray Conference}
\CopyRight{\copyright 2003 by Universal Academy Press, Inc.}
%\tableofcontents
\pagenumbering{arabic}

\chapter{Atmospheric Monitoring for the Pierre Auger Fluorescence Detector}

\author{Miguel A. Mostaf\'a$^1$ for the Pierre Auger Collaboration$^2$ \\
{\it (1) University of New Mexico, Albuquerque, NM 87131\\
(2) Observatorio Pierre Auger, %Av. San Mart\'{\i}n Norte 304, 
Malarg\"ue, (5613) Mendoza, Argentina}}

\section*{Abstract}

Major uncertainties 
in the fluorescence energy measurement 
come from the precision of various atmospheric transmission,
air Cerenkov subtraction, light multiple-scattering and cloud corrections.
The Auger program for atmospheric monitoring, 
designed to measure these corrections and to minimize these uncertainties,
is summarized in this paper.

\section{Introduction}

A reliable air shower energy scale and energy resolution is central to the
mission of the Pierre Auger ultra-high energy cosmic ray experiment.
The Auger Southern Observatory is able to measure the energy of the
incoming cosmic rays using fluorescence detectors, and thus calibrate
the energy inferred from the surface detectors.
Uncertainties in the fluorescence energy measurement must therefore
be well understood.

For fluorescence experiments, the atmosphere not only acts as the showering medium for the primary
cosmic ray, but it is also an essential part of the readout system.
Thus, the atmosphere must be calibrated,
the calibration monitored with time,
and then %the atmospheric calibration 
input to the analysis of the fluorescence data.
%
%Especially, light scattering and absorption by aerosols, such as smoke, dust or fog, are very complicated
%as well as time-dependent.
%Thus, understanding the atmosphere is a key issue on shower energy and flux calculations.
Auger atmospheric monitoring systems %to characterize the atmosphere using
%The calorimetric energies of air showers should be measured carefully by characterizing light
%propagation in the atmosphere with various atmospheric monitoring 
include
%tools such as
LIDARs, cloud monitors, horizontal attenuation monitors, phase function monitors,
meteorological stations, and radiosondes.
The main systems deployed on-site are described in this paper.
%, and preliminary results are discussed.

\section{Monitoring systems}

\subsection{LIDAR Stations}

The primary source of atmospheric information for the Auger fluorescence detectors
derives from steerable backscattered LIDAR systems.
As these are located at each fluorescence site,
they are able to probe directly the shower-detector atmospheric path.
These systems %have to perform two main tasks.
%The first is to 
are used to
make routine surveys of the vertical profile of aerosols around the local fluorescence
detector.
Immediately following a promising shower event,
they scan the atmosphere %in the direction of 
%several times per night,
%The purpose of these last measurements is to 
looking for possible scattering inhomogeneities 
%within the shower-detector plane.
between the shower and the fluorescence detector.

Each LIDAR station  % Description
consists of a pulsed laser beam at 355 nm, a receiver telescope with three mirrors,
each with a gated, high-speed photon detector [2].
%The laser is a flush lamp pumped, Q-switched, and water cooled Nd:YAG type.
% HOW MUCH DETAIL SHOULD I INCLUDE???
The receiver measures the backscattered photons as a function of time,
\textit{i.e.} intensity of photons versus distance to the point where the light backscattered.
The aerosol optical depth as a function of height is inferred from these 
measurements~[3].

The first LIDAR telescope was installed at the Los Leones site in February, 2002.
This station has been operational since April, 2002. 
Figure~(\ref{figura}a) shows a typical measurement of the aerosol vertical profile at 
the Los Leones site.
Several inversion methods are currently being compared.
%and used to validate the system and evaluate data quality.
The second telescope was installed at the Coihueco site in May, 2003.

\subsection{Horizontal Attenuation Length Monitor}

This system %routinely and automatically 
records the %combined Rayleigh and Mie 
horizontal attenuation length
%$\Lambda(\lambda) = [1/\Lambda^{m}(\lambda) + 1/\Lambda^{a}(\lambda)]^{-1}$,
at several wavelengths within and near to the acceptance of the fluorescence detectors.
For the Auger Southern Observatory three systems will monitor three different light paths across the site.
%They will provide a single parameter by which the atmospheric clarity can be assessed,
%and also information on instrument related systematic uncertainties.

Each system includes a stable DC light (source) 
and a UV sensitive CCD camera (photometer).
%Mercury vapor lamps are used as they have strong emission lines in the UV.
Measurements are made through interference filters at 365~nm, 405 nm, 436 nm, and 542 nm.
The total horizontal attenuation length is calculated
from the ratio of flux measured %at the photometer,
%which is normally positioned 
at a large distance ($\sim 50$ km) from the light source 
to the flux measured
at a small distance
(calibration point).

The first system was installed in May, 2001.
It produces three sets of measurements per night.
%with five images per wavelength, and a couple of system background and
%sky noise measurements.
%This prototype has demonstrated that the system can operate automatically for long periods of time,
%and has shown that data \textit{make sense...}. 
%Reconstruction algorithms were tested on almost one year of real data,
%and applied to extract daily horizontal attenuation lengths.

\subsection{Aerosol Phase Function Monitor}

The goal of this system is to measure the normalized aerosol differential scattering cross section 
%$\frac{1}{\sigma^{a}}\frac{d\sigma^{a}}{d\Omega}$, 
as a function of the scattering angle from the initial light direction.
%In the constant composition, 1-dimensional model for aerosols it is sufficient to measure the aerosol phase function
%at the altitude of the fluorescence detectors.
The measurement is made using a horizontal pulsed light beam  directed across the field of view of one of the
fluorescence detectors [6].
%As each fluorescence site views $180^{o}$ in azimuth, even a fixed direction light beam provides a monitor 
%of the aerosol phase function over most of the range of scattering angles
%(between $20^{o}$ and $160^{o}$).

The procedure to measure the phase function employs three xenon flash tube sources
which emit $1~\mu s$ light pulses covering a range of wavelengths [5].
%The wavelength intervals are 
%selected using optical filters. 
%and a small telecope is used to collimate the light.
The distance between the light source and the fluorescence building is $\sim 1.3$ km,
which is much less than the total atmospheric attenuation length.
This geometry requires only small corrections for transmission and multiple scattering,
and allows the fluorescence detector to measure the light scattered out of the beam over the desired range
of scattering angles
(between $20^{o}$ and $160^{o}$).

The first aerosol phase function light sources were installed in March, 2003. 
%and it is ready to start producing data at the begining of the next fluorescence data taking period.

\subsection{Cloud Cameras}

These systems allow cloud detection over the Auger array.
Clouds have the potential to modify the fluorescence measurements in an
unpredictable way.
%which is a crucial factor in determining the collecting area available while
%fluorescence observations are in progress.

The detection of clouds has been proved possible using their strong infra-red emission
against a much weaker clear sky background [1].
The detector is based on 
the Raytheon Control-IR 2000B 
%an IR
digital camera with a spectral range
%The detector is based on a digital IR camera with a spectral range
between $7$ and $14~\mu$m,
and a field of view of $45^{o}$ x $36^{o}$.
%, 12 bit resolution, and a maximum frame rate of 30 Hz.
Operation modes include: 
scan the field of view of the fluorescence telescopes, % every five minutes,
%produce a mosaic picture of the full sky, % every fifteen minutes,
produce a full-sky image, % every fifteen minutes,
% Take out "receiving"?
and take a picture of a given direction after %receiving 
a trigger from the fluorescence detector.
%(with some angular velocity cut to reduce the rate to 20 images per night).

\begin{figure}[t]
  \begin{center}
    \fbox{\includegraphics[height=14.0pc]{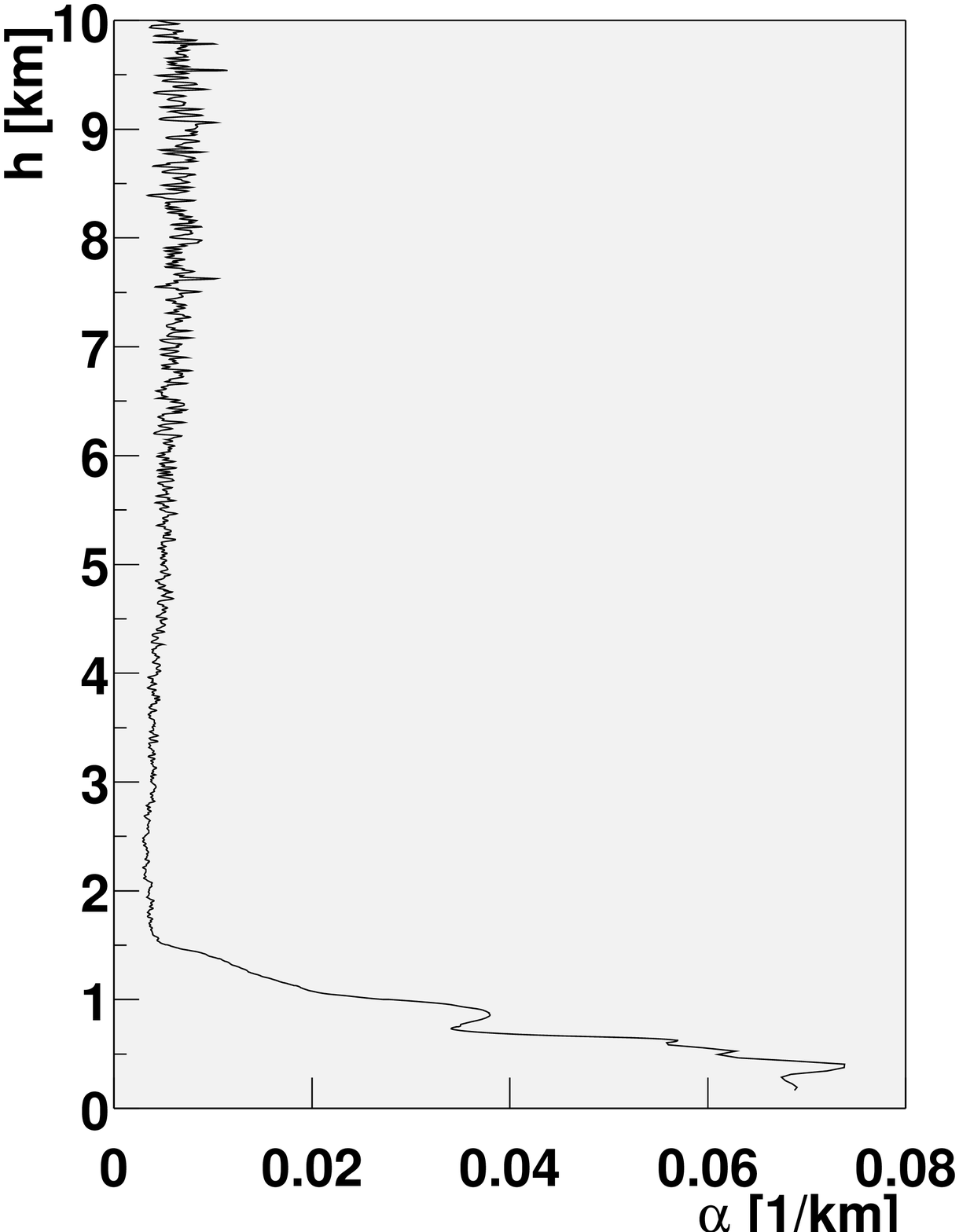}}%%%%% ``13.5pc'' is just the example.
\hfill
    \fbox{\includegraphics[height=14.0pc]{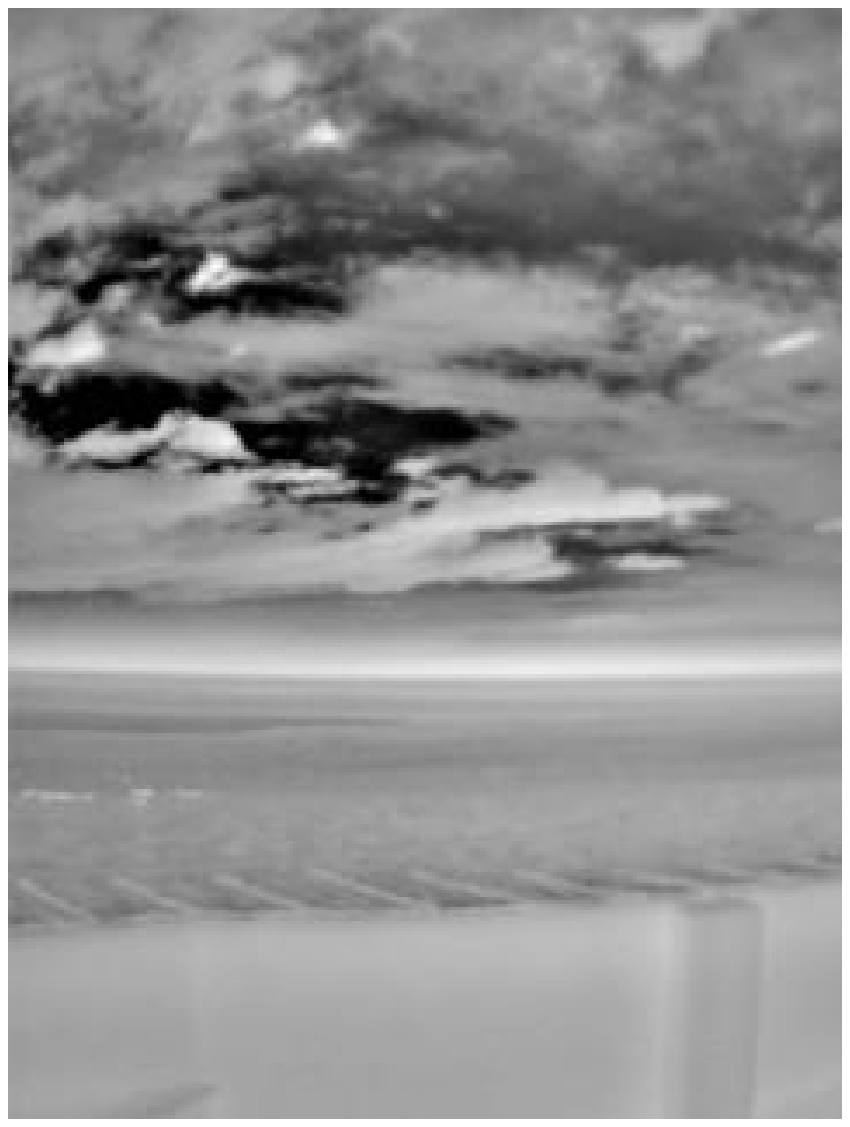}}%%%%% ``13.5pc'' is just the example.
\hfill
    \fbox{\includegraphics[height=14.0pc]{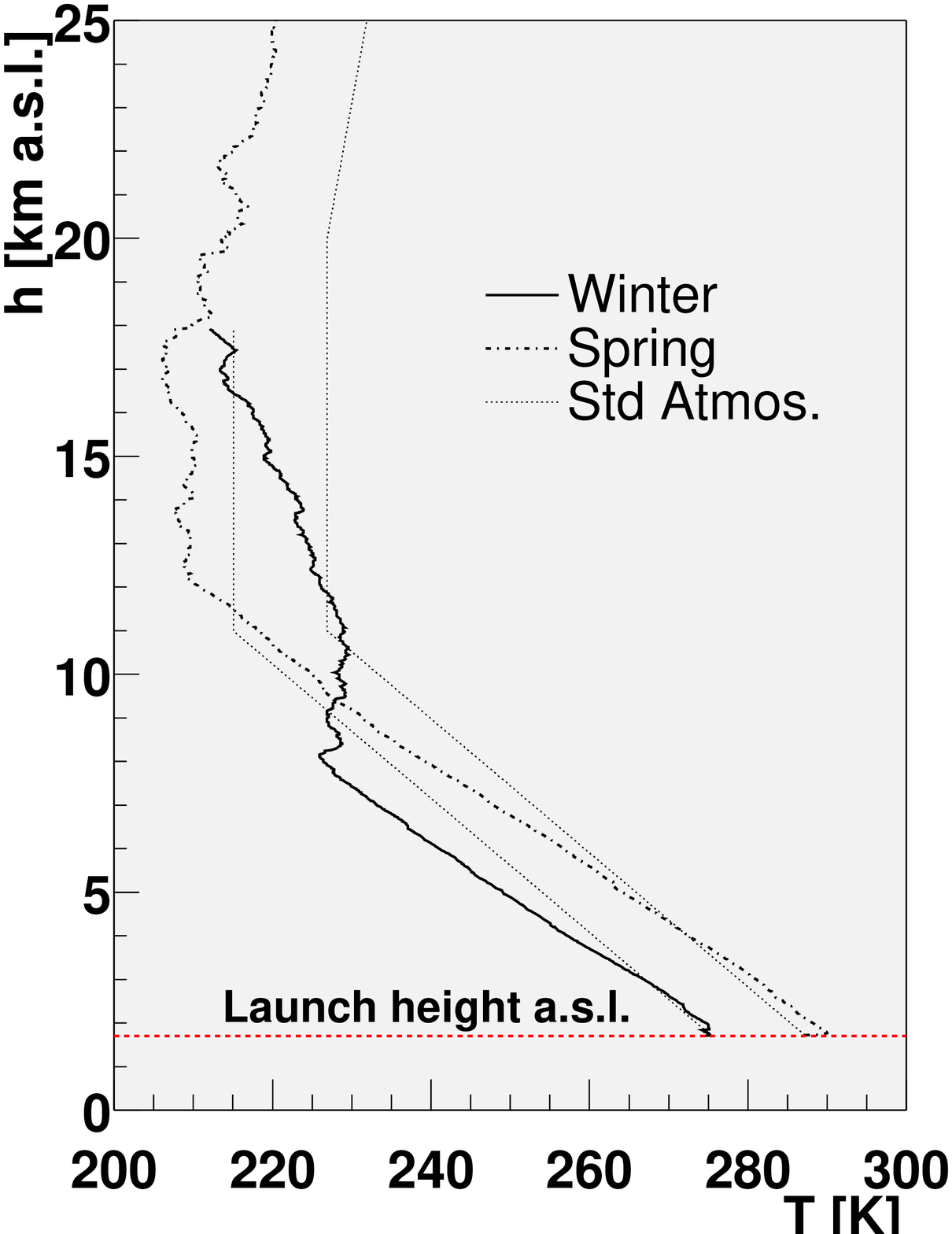}}\\%%%%% ``13.5pc'' is just the example.
%(a)\hfill(b)\hfill(c)
%(a)\hspace{10.125pc}(b)\hspace{10.125pc}(c)
\hspace*{5.0625pc}(a)\hfill(b)\hfill(c)\hspace*{5.0625pc}
  \end{center}
  \vspace{-0.5pc}
  \caption{\label{figura}
(a) Aerosol extinction coefficient as a function of height measured with the LIDAR system at the Los Leones site.
(b) Sky picture of the field of view of one of the telescopes at the Los Leones site.
%This sample image was taken with the fixed cloud camera prototype installed in July, 2002.
(c) Temperature profiles measured with radiosondes in two different seasons over the Coihueco site.
}
\end{figure}

The prototype, a fixed IR camera, 
and was installed in July, 2002.
It
%This system %reliably operates since then for 24 hours a day, and it 
produces 24 representative images per night. 
A sample image %of the Los Leones fluorescence site
is shown in figure~(\ref{figura}b).
Two scanning systems will be installed this year.
%and their control will be incorporated to the on-line data acquisition system.
Work is in progress to improve methods for determining cloud edges,
and increasing the sensitivity to high thin clouds.

\subsection{Meteorological Stations and Radiosondes}

There are three weather stations operating %identified by location
at the Los Leones, Coihueco, and a central site since 2001.
They provide daily ground measurements of
 temperature,
relative humidity,
wind speed
and direction,
solar radiance,
and %barometric 
pressure.

Meteorological radiosondes are used to
measure temperature and pressure as a function of height,
and to %determine air density profiles.
%They monitor atmospheric changes over the different seasons, and 
cross check the results of the standard profiles based on 
measurements at
ground %values.
level.

These radiosondes are standard meteorological instruments for measuring
pressure, temperature, and humidity. % [4].
They are launched using helium filled balloons, and
raw data,
including GPS information,
are received by a radio ground station. % on a carrier frequency of 404 MHz with a bandwidth of 15 kHz.

Measurement campaigns have been performed since August, 2002
during the local winter, spring, and summer.
%for a total of ... launches.
Balloons are launched from different starting positions at different times during 
the day and night.
%On average, the rate of climb was of the order of 200 m/min,
%and the nominal burst height was 15 km.
Figure~(\ref{figura}c) shows measured temperature profiles for winter and spring over the Coihueco site.

The air density as a function of height is calculated based on these data.
Results are compared to standard profiles, 
and used to study the 
effects on the longitudinal shower development and fluorescence yield.
A detailed description of the radiosondes, the measurements, and the analysis can be found in Ref. [4].
Work in progress concentrates on the development of a corrected standard profile to describe the data based only
on ground %values. % of pressure and temperature.
level meteorological measurements.

\section{Cross Checks} % (within Conclusions?)

To minimize systematic uncertainties,
all the atmospheric measurements are made in at least two independent ways.
The horizontal attenuation length is compared with the results of horizontal LIDAR measurements.
The aerosol optical depth can be cross checked using dedicated Raman LIDAR,
and also with %roving laser systems.
a central vertical laser.
The side scattered light from a central steerable laser observed at the fluorescence detectors provides
an essential cross check of the aerosol model.
In addition, LIDAR routine scans can be used to determine cloud base height, which combined with
the cloud camera information, produce a detailed 3D picture of the clouds.

%Lidar vs. HAM; + LIDAR for clouds in 3D; + 
%Atmospheric changes affecting the total optical depth.
%Other systems are also used, like roving lasers, and other laser-based monitors to cross checks ...

\section{Conclusions}

The main components of the Auger atmospheric monitoring program have already been deployed
on-site. 
Atmospheric data are being analyzed and the first atmospheric data bases are being constructed.
The Auger monitoring goal is to limit the atmospheric contributions to the shower energy uncertainty 
to $\sim 10 \%$.

%\textit{A bunch of stuff is coming along!!!}

%\section{List of Symbols/Nomenclature}
%
%  \begin{tabbing}
%     LIDAR = Light detection and ranging    \=       \kill
%     LIDAR = Light detection and ranging    \>       \\
%     US-StdA = US Standard Atmosphere       \>  p=Pressure, Kg/m$^{2}$    \\
%     S=Source Term                           \>  T=Temperature, $^{\circ}$C
%\\
%     t=Time, sec                            \>p'=Effective Pressure, Kg/m$^{2}$
%  \end{tabbing}

\section{References}

%\vspace{\baselineskip}
%\re
\noindent
1.\ Clay R. W., \textit{et al.}\ 1998, Pub. Astron. Soc. Aust. \textbf{15}, 334
\re
2.\ Filip\v{c}i\v{c} A., \textit{et al.}\ 2001, Proc. of $27^{th}$ ICRC, \textbf{2}, 784
%International Cosmic Ray Conference, \textbf{2}, 784
\re
3.\ Filip\v{c}i\v{c} A., \textit{et al.}\ 2003, Astroparticle Physics \textbf{18}, pp 501-512
\re
4.\ Keilhauer B., \textit{et al.}\ 2003,  Proc. of $28^{th}$ ICRC, and Auger Note GAP-2003-009
\re
5.\ Matthews J. A. J., and Mostaf\'a M. A.\ 2003,  Proc. of $28^{th}$ ICRC
\re
6.\ Tessier T., \textit{et al.}\ 1999, Proc. of $26^{th}$ ICRC, \textbf{5}, 408
%International Cosmic Ray Conference, \textbf{5}, 408

\endofpaper
\end{document}